\documentclass[12pt]{article}
 \usepackage{graphicx}
 \usepackage[cp1251]{inputenc}

\begin{document}

 \bigskip
 \bigskip
 \centerline {\Large\bf Investigation of the Galactic Bar Based on Photometry}
 \centerline {\Large\bf and Stellar Proper Motions}
 \bigskip
 \centerline{V. V. Bobylev$^{1, 2}$, A. V. Mosenkov$^{1, 2}$, A. T. Bajkova$^1$, and G. A. Gontcharov$^1$}

 \bigskip
 \centerline{\small $^{1}$\it Pulkovo Astronomical Observatory, Russian Academy of Sciences,}
 \centerline{\small \it Pulkovskoe sh. 65, St. Petersburg, 196140 Russia}
 \centerline{\small $^{2}$\it Sobolev Astronomical Institute, St. Petersburg State University,}
 \centerline{\small \it Universitetskii pr. 28, Petrodvorets, 198504 Russia}
 \bigskip
 \bigskip

\noindent {\bf Abstract} -- A new method for selecting stars in
the Galactic bar based on 2MASS infrared photometry in combination
with stellar proper motions from the Kharkiv XPM catalogue has
been implemented. In accordance with this method, red clump and
red giant branch stars are preselected on the color -- magnitude
diagram and their photometric distances are calculated. Since the
stellar proper motions are indicators of a larger velocity
dispersion toward the bar and the spiral arms compared to the
stars with circular orbits, applying the constraints on the proper
motions of the preselected stars that take into account the
Galactic rotation has allowed the background stars to be
eliminated. Based on a joint analysis of the velocities of the
selected stars and their distribution on the Galactic plane, we
have confidently identified the segment of the Galactic bar
nearest to the Sun with an orientation of 20$^\circ$--25$^\circ$
with respect to the Galactic center -- Sun direction and a
semimajor axis of no more than 3 kpc.

\bigskip\noindent {\bf DOI:} 10.1134/S1063773714030037


\newpage
\section*{INTRODUCTION}

The first evidence for the presence of a central bar in the Galaxy
was probably provided by de Vaucouleurs (1970). Subsequently,
observational proof of the existence of a bar was obtained in
several ways. These include a study of the gas kinematics (Binney
et al. 1991), analysis of the surface brightness (Blitz and
Spergel 1991), star counts (Nakada et al. 1991; Stanek et al.
1994), and experiments on searching for microlensing (Udalski et
al. 1994).

The appearance of infrared photometry for hundreds of millions of
stars led to a considerable success in studying the Galactic bar.
Analysis of the red giant clump from the 2MASS (Babusiaux and
Gilmore 2005), OGLE-II (Rattenbury et al. 2007), and OGLE-III
(Nataf et al. 2013) catalogues showed that the bar is oriented
with respect to the Galactic center.Sun direction at an angle of
15$^\circ$--45$^\circ$, its radius is 3--4 kpc, and the ratio of
its axes $(x_b : y_b : z_b)$ is approximately 10 : 3.6 : 2.7
(Rattenbury et al. 2007). At present, the bar orientation
parameters and sizes are being debated. Most authors incline to
the model of a short bar ($R_b\approx3$~kpc) oriented at an angle
of about 20$^\circ$. Other authors talk about a long bar
($R_b\approx4$~4 kpc) oriented at an angle of about 44$^\circ$
(Benjamin et al. 2005; L$\acute{o}$pez -- Corredoira et al. 2007).

We are going to use red clump and red giant branch stars as
distance indicators. The method for selecting such stars was
implemented by Gontcharov (2008, 2011) using the2MASS (Skrutskie
et al. 2006) and Tycho-2 (Hog et al. 2000) catalogues as an
example.

It is important to note that the motions of stars belonging to the
bar have significant deviations from circular orbits. If we take
many stars lying on the same line of sight (toward the bar) but
located at different distances (at both near and far ends of the
bar), then we will observe a considerable velocity dispersion.

Simulations of the bar kinematics show that there is a peak in the
dispersions of both stellar line-of-sight velocities (Zhao 1996;
Wang et al. 2012) and proper motions (Zhao 1996) in the direction
$l=0^\circ$. We are going to use this fact for a more reliable
selection of stars belonging to the Galactic bar. The
line-of-sight velocities of stars are of greatest interest for
such a task, because their values and errors do not depend on
heliocentric distance. At present, however, there is no sufficient
number of line-of-sight velocity measurements for the stars we
need. Therefore, we are going to use the proper motions of stars.

The Kharkiv XPM catalogue (Fedorov et al. 2009, 2011), which
contains the proper motions of $\sim$314 million stars and
incorporates 2MASS infrared photometry, is quite suitable for
these purposes. The fact that the XPM catalogue is an independent
realization of the inertial reference frame is of particular
interest. About 1.5 million galaxies were used to absolutize the
stellar proper motions.

The spiral structure of the Galaxy is closely related to the bar.
At present, however, there is no unequivocal answer even to the
question about the number of spiral arms in the Galaxy. Analysis
of the spatial distribution of young Galactic objects (young
stars, star-forming regions, open star clusters, and hydrogen
clouds) shows that two-, three-, and four-armed patters are
possible (Russeil 2003; Vall$\acute{e}$e 2008; Hou et al. 2009;
Efremov 2011; Francis and Anderson 2012). More complex models are
also known, for example, the kinematic model by L$\acute{e}$pine
et al. (2001), where two- and four-armed patterns combine.
According to Englmaier et al. (2011), the HI distribution in the
Galaxy suggests that a two-armed pattern is possible in the inner
part $(R < R_0)$ of the Milky Way Galaxy, which breaks up into a
four-armed one in the outer part $(R > R_0)$. Note also the
spiral–ring model of the Galaxy (Mel’nik and Rautiainen 2009),
which includes two outer rings elongated perpendicular and
parallel to the central Galactic bar, an inner ring elongated
parallel to the bar, and two small fragments of spiral arms.

The 3-kpc arm segment is most closely related to the bar. Its
peculiarity is a considerable radial (recession from the Galactic
center) velocity of about 50 km s$^{-1}$ (Burton 1988; Sanna et
al. 2009). The 3- kpc arm segment located behind the bar, in the
region of positive Galactic longitudes, is also seen at present
(Dame and Thaddeus 2008). Thus, the 3-kpc spiral arm segments
contribute significantly to the velocity dispersion toward the
bar. Therefore, we separate the stars into two groups. The first
group includes the stars of the bar and spirals; the second group
includes all of the rest (disk stars with nearly circular orbits
and “background stars”).

This paper is devoted to the selection of stars belonging to the
Galactic bar and the spiral arms in the central part of the Galaxy
and to their analysis to determine the bar geometry. We estimate
the photometric distances of red clump and red giant branch stars
from infrared photometry. We use the stellar proper motions as
indicators of the velocity dispersion toward the bar and the
spiral arms. As a result, this allows the stars whose orbits
differ significantly from circular ones to be analyzed. Naturally,
such an approach required a careful allowance for the Galactic
rotation.

\section*{METHODS}
\subsection*{\it The Red Giant Clump and Red Giant Branch Stars}
To identify probable stars of the Galactic bar, we used the
following selection criteria from the Kharkiv XPM catalogue and
the 2MASS catalogue:
\begin{itemize}
 \item[---] $|l|<90\,^{\circ}$, $|b|<10\,^{\circ}$\,,
 \item[---] Reliable 2MASS photometry: $K_{s}<14^{m} $, photometric
errors $\sigma(J)<0.^{m}05$, $\sigma(H)<0.^{m}05$,
$\sigma(K_{s})<0.^{m}05$.
\end{itemize}

The initial sample consisted of 30 million stars. Using the 3D
extinction map in the Galaxy (Marshall et al. 2006), we
constructed a linear fit to the interstellar extinction as a
function of distance:
\begin{equation}
 A_\mathrm{Ks} \sim k\cdot r\, ,
 \label{ext}
\end{equation}
where k is the coefficient determined for each
$0.25\,^{\circ}\times0.25\,^{\circ}$ region of the celestial
sphere (Fig. 1). The distance to a star $r$ was determined from
the well-known formula
\begin{equation}
 \lg r=1+0.2(K_s-M_{K_s}-A_{K_s}(r)) \,.
 \label{dist-0}
\end{equation}
For red giant clump stars, we took $M_{K_s}=1.52^{m}$ (Gontcharov
2008). The true color index was found as follows:
\begin{equation}
(J-K_{s})_{0}=(J-K_{s})-E(J-K_{s}) \, ,
 \label{dist}
\end{equation}
where we applied the following relation to estimate the color
excess, in accordance with the extinction law from Draine (2003):
\begin{equation}
 E(J-K_{s})=A_{K_{s}}(r)/0.67.
 \label{E}
\end{equation}
Finally, to find the absolute magnitudes of the red giant branch
stars, we used the following polynomial that we obtained from the
sample of red giant branch stars by Gontcharov (2011):
\begin{equation}
 M_{K_{s}}=\varphi((J-K_{s})_{0}),
 \label{Mabs}
\end{equation}
where $\varphi(x) = 3.5373\,x^2-14.745\,x+7.13$.

Thus, the distances to the red giant branch stars were found by
solving the equation
\begin{equation}
 \lg r = 1+0.2\,[ K_{s}-\varphi(J-K_{s}-k\cdot r/0.67)-k\cdot r]\,.
 \label{dist1}
\end{equation}
To find the roots of this equation, we applied the standard
bisector method.

For the red clump giants, the distances can be calculated from a
simpler formula:
\begin{equation}
 \lg r = 1+0.2\,( K_{s}+1.52-k\cdot r).
 \label{dist2}
\end{equation}
For each star from the sample, we determine the true color index
$(J-K_s)_0$ by applying a linear interstellar extinction law with
coefficient $k$ for the field into which a given star falls. Next,
we construct a color–-magnitude diagram for each field.

The samples of red clump and red giant branch stars were produced
as follows. Having analyzed the diagram for each field, we put the
stars with a true color index $(J-K_s)_0>1^{m}.0$ into the sample
of red giant branch stars. The condition
$0^{m}.55<(J-K_s)_0<0^{m}.9$ should be met for the red clump
stars. Next, we selected only the stars with $K_s<12^m$ to rule
out the falling of many main-sequence stars into our sample.

Figure 2 presents an example of such a diagram for a specific sky
region.

\begin{figure}[t]
{\begin{center}
 \includegraphics[width=100mm]{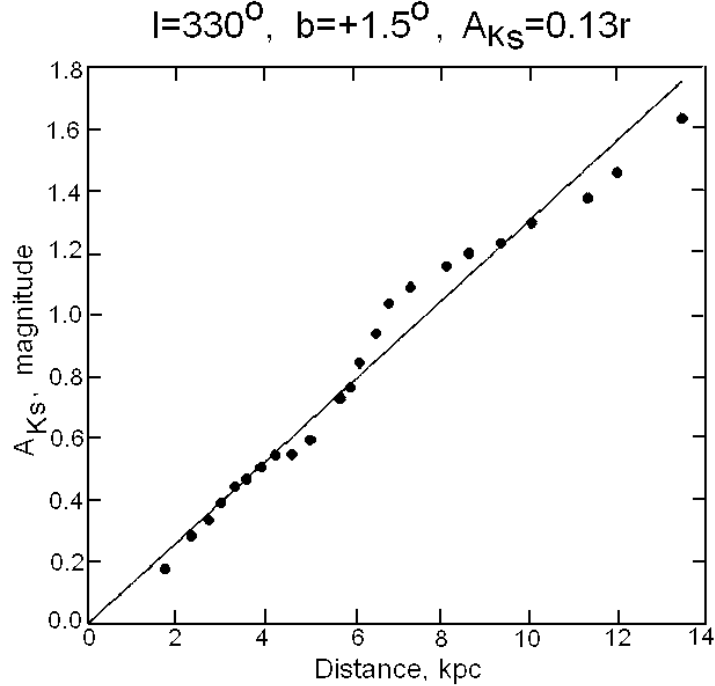}
 \caption{
Example of constructing a linear fit to the law of interstellar
extinction as a function of distance for a given sky region.}
 \label{ext_example}
\end{center}}
\end{figure}

\begin{figure}[t]
{\begin{center}
 \includegraphics[width=140mm]{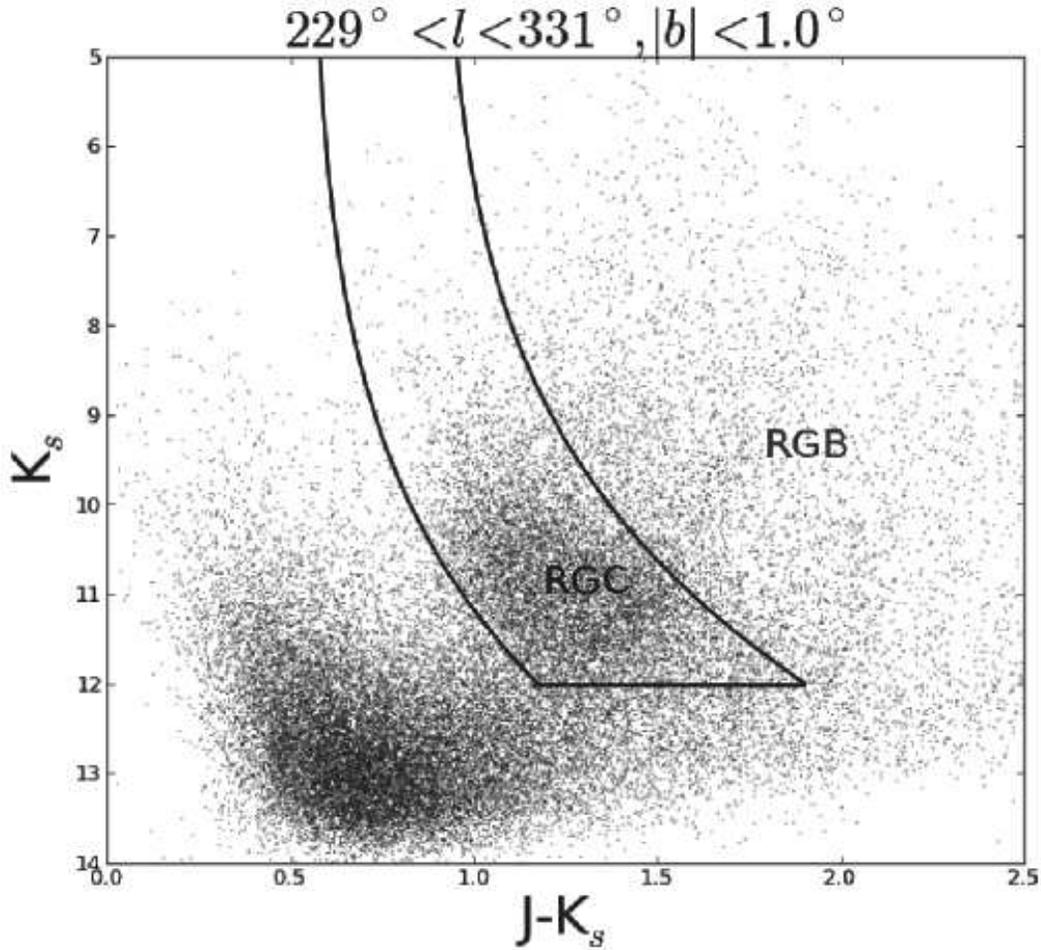}
 \caption{
Example of identifying the red giant clump (RGC) and the red giant
branch (RGB) on the color–magnitude diagram for a selected sky
region.}
 \label{DGR_example}
\end{center}}
\end{figure}

Thus, we obtained two samples of stars: red giant branch
($\approx$1.3 million) and red clump ($\approx$10 million) stars
with their photometric distances derived from Eqs. (6) and (7),
respectively.

\subsection*{\it The Application of Stellar Proper Motions}

We used the stellar proper motions from the Kharkiv XPM catalogue,
which contains the positions and absolute proper motions of $\sim
314$ million stars. The stars cover the entire celestial sphere
and have magnitudes in the range $10^m<B<22^m$. For the
overwhelming majority of stars, the XPM catalogue contains their
infrared $JHK$ magnitudes from the 2MASS catalogue. According to
the estimate by Fedorov et al. (2011), the random errors of the
XPM stellar proper motions are 3--8 and 5--10 mas yr.1 for the
northern and southern skies, respectively.

The parameters of the Galactic rotation curve containing six terms
of the Taylor expansion of the angular velocity of Galactic
rotation $\Omega$ at the Galactocentric distance of the Sun
$R_0=7.5$~kpc were found by Bobylev et al. (2008). Data on
hydrogen clouds at tangential points (Clemens 1985), on massive
star forming regions (Russeil 2003), and the velocities of young
open star clusters were used for this purpose. The more recent
value of R0 is $8.0\pm0.4$ kpc (Foster and Cooper 2010).
Therefore, we redetermined the parameters of the Galactic rotation
curve based on the same sample but for $R_0=8$~kpc:
 \begin{equation}
  \begin{array}{lll}
 \Omega_0  =  -27.4\pm 0.6~\hbox {km s$^{-1}$ kpc$^{-1}$},  \\
 \Omega^1_0= ~~3.80\pm0.07~\hbox {km s$^{-1}$ kpc$^{-2}$},  \\
 \Omega^2_0= -0.650\pm0.065~\hbox {km s$^{-1}$ kpc$^{-3}$}, \\
 \Omega^3_0=~~0.142\pm0.036~\hbox {km s$^{-1}$ kpc$^{-4}$}, \\
 \Omega^4_0= -0.246\pm0.034~\hbox {km s$^{-1}$ kpc$^{-5}$}, \\
 \Omega^5_0=~~0.109\pm0.020~\hbox {km s$^{-1}$ kpc$^{-6}$}.
   \label{Omega}
  \end{array}
 \end{equation}
The Galactic rotation curve $V_{rot}=R\Omega$ constructed with
parameters (8) is shown in Fig. 3.

\begin{figure}[t]
{\begin{center}
 \includegraphics[width=140mm]{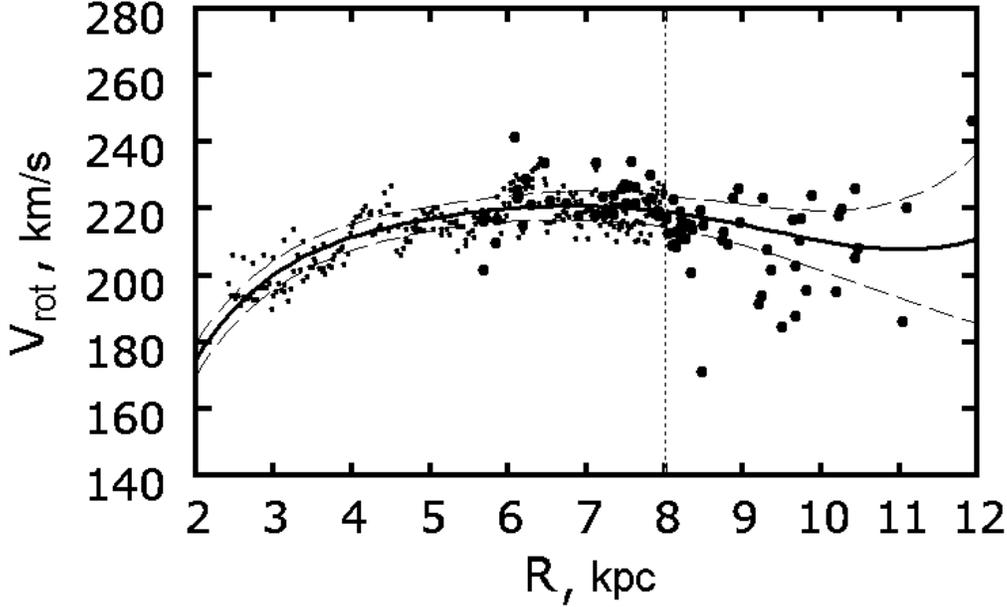}
 \caption{
Galactic rotation curve (thick line) constructed with parameters
(8), the dotted lines mark the 1. confidence region, the vertical
line marks the Sun’s position, the black and gray circles indicate
the HI and HII velocities, respectively. }
 \label{f3}
\end{center}}
\end{figure}

Our calculations of the mean values based on a sample of distant
stars showed the mean dispersion of the XPM stellar proper motions
to be $\sigma_\mu\approx8$ mas yr$^{-1}$ in each coordinate:
 \begin{equation}
  \begin{array}{lll}
 \overline{\mu_\alpha\cos\delta}=-2.1\pm7.6~\hbox{ mas yr$^{-1}$} \\
 \overline{\mu_\delta}= -4.4\pm7.7~\hbox{ mas yr$^{-1}$}.
   \label{mu}
  \end{array}
 \end{equation}
A linear velocity of 303 km s$^{-1}$ for the distance $r=8$ kpc
corresponds to $\sigma_\mu=8$mas yr$^{-1}$. The means (9) reflect
the motion of the Sun around the Galactic center. They should be
compared with the values of $\mu_\alpha\cos\delta=-3.151\pm0.018$
mas yr$^{-1}$ and $\mu_\delta= -5.547\pm0.026$ mas yr$^{-1}$
obtained by Reid and Brunthaler (2004) by VLBI based on an
eight-year-long monitoring of the positions of the radio source
Sgr~A relative to two quasars. Attempts to obtain the limiting
mean proper motions from the most distant stars of the XPM
catalogue or only from the stars near the Galactic center
direction showed that the results differ only slightly from the
means (9). We may conclude that the means (9) differ from the
results by Reid and Brunthaler (2004) in each of the coordinates
by $\approx$ mas yr$^{-1}$. Such a difference may be related to
peculiarities of the absolutization of stellar proper motions in
the XPM catalogue. In spite of the fact that data on tens of
thousands of extragalactic sources were used for this procedure,
there were no such data near the Galactic plane. Therefore, an
extrapolation procedure was applied here (Fedorov et al. 2009). On
the other hand, as can be seen from Fig. 5, the mean radius of the
sample of stars is about 5 kpc; therefore, the means (9) are
slightly smaller than the expected values for a distance of 8 kpc.

\begin{figure}[p]
{\begin{center}
 \includegraphics[width=140mm]{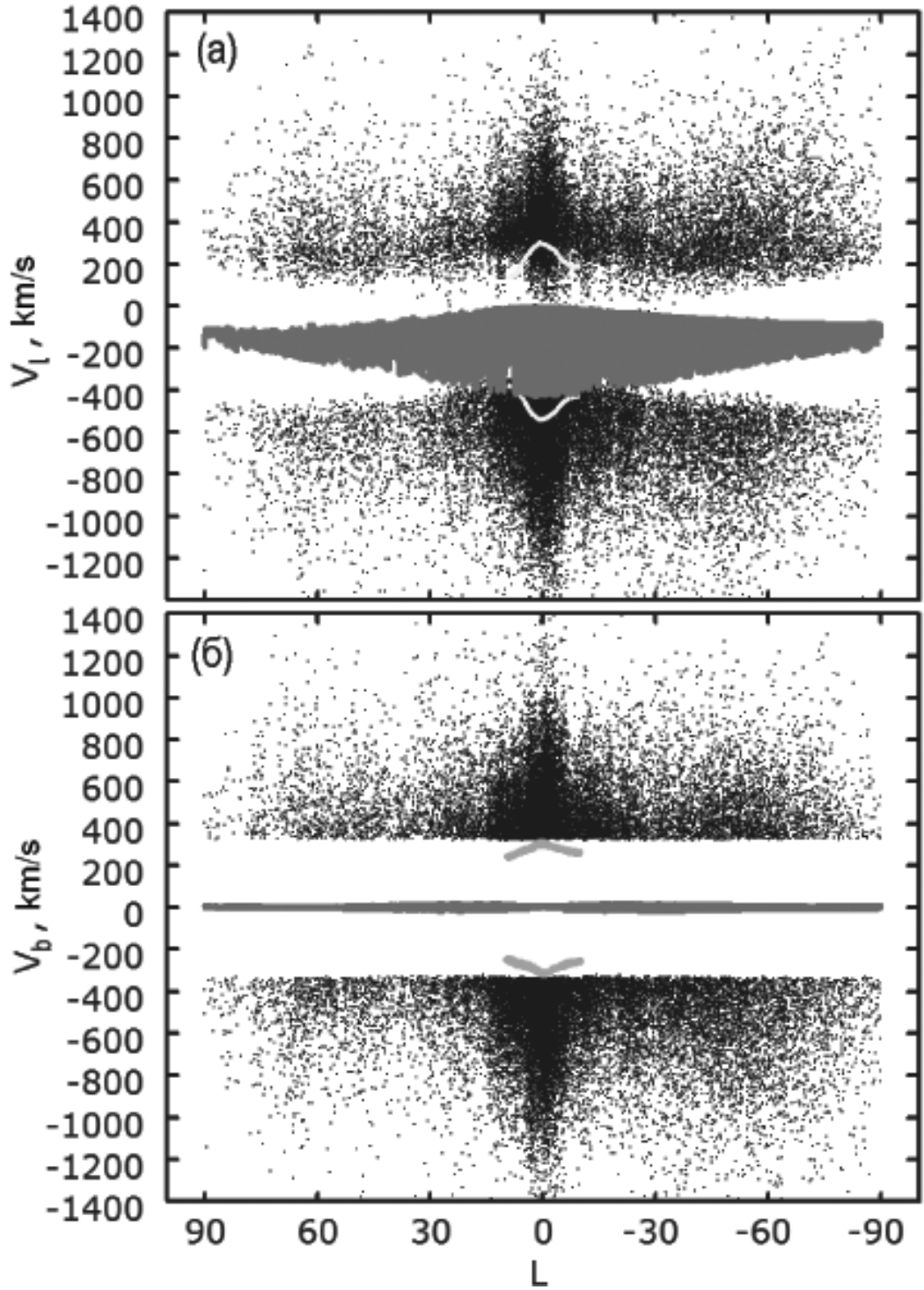}
 \caption{
Velocities of red giant branch stars $V_l$ (a) and $V_b$ (b)
versus Galactic longitude $l$; additional explanations are given
in the text. }
 \label{f4}
\end{center}}
\end{figure}

\begin{figure}[p]
{\begin{center}
 \includegraphics[width=140mm]{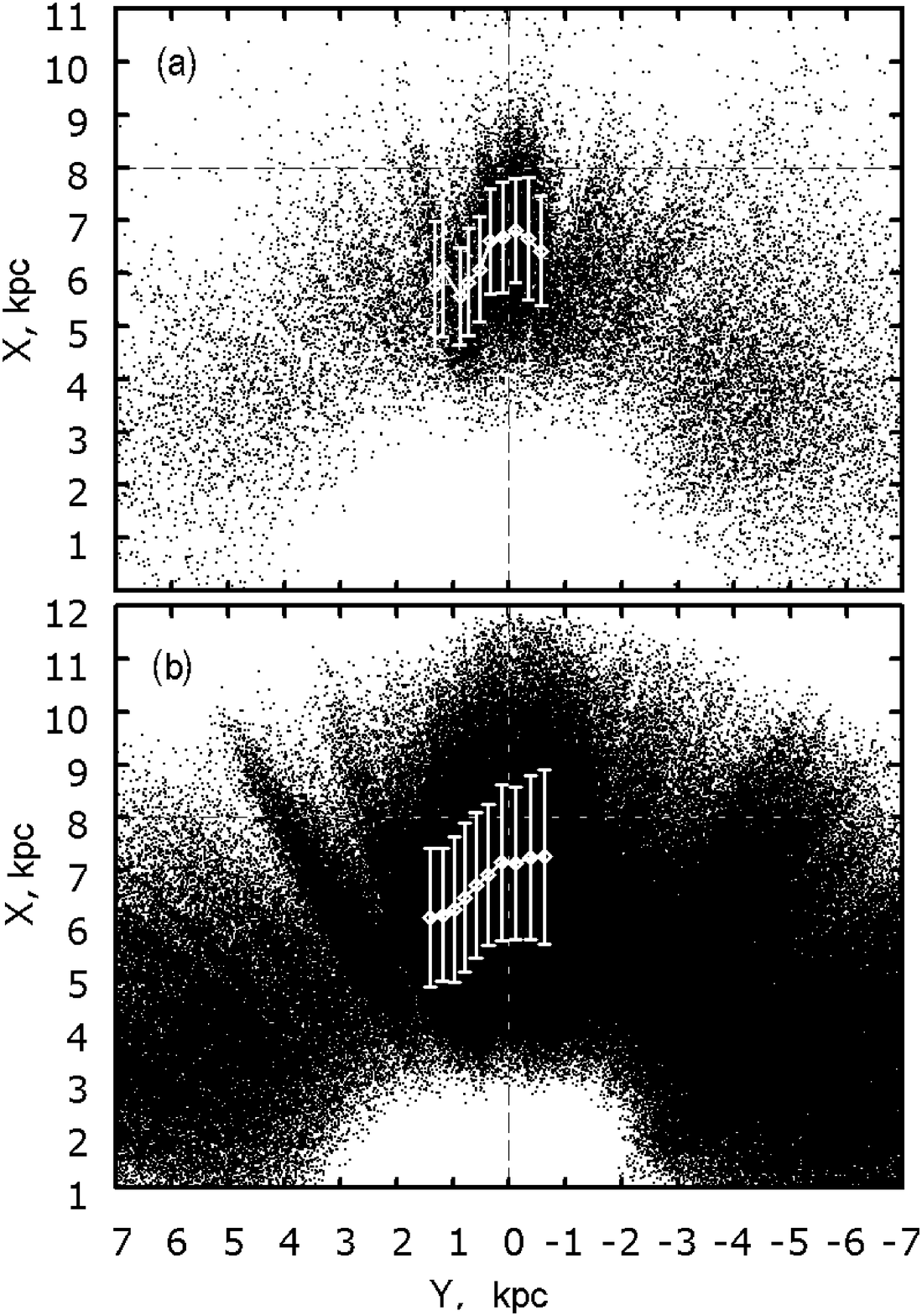}
 \caption{
Distributions of red giant branch (a) and red clump (b) stars in
projection onto the Galactic plane, the Sun is at the coordinate
origin, the presumed Galactic center for $R_0=8$~ kpc is located
at the intersection of the dashed lines. }
 \label{f5}
\end{center}}
\end{figure}

The application of stellar proper motions consists in the
following.

(1) We take the stars close to the expected motion of the Galactic
center: $|\mu_\alpha\cos\delta-(-3.2)|<30$~mas yr$^{-1}$ and
$|\mu_\delta-(-5.5)|<30$~mas yr$^{-1}$, where the values from Reid
and Brunthaler (2004) were taken as the means.

(2)We take the stars whose velocities deviate significantly, by
$>$330~ km s$^{-1}$, from the circular rotation velocity,
$\approx~220$~km s$^{-1}$. The circular velocities are calculated
from the Galactic rotation curve with parameters (8). This
condition allows the stars with nearly circular orbits and
foreground stars to be rejected.

(3) We use the constraint on the limiting radius of the sample r <
12 kpc. The upper limit of the observed velocities for the
distance $r=12$~kpc and $\sigma_\mu=8$~mas yr$^{-1}$ is estimated
to be $4.74 r 3\sigma_\mu=1642$~km s$^{-1}$. As can be seen from
Fig. 4, there are virtually no stars with such large velocities in
our sample.

(4) At low Galactic latitudes, $|b|<2^\circ$, there is an
indistinct clump of stars toward the Galactic center. The
extinction is very large here and the stars of the bar are very
difficult to identify. Therefore, we do not use this region. As a
result, we obtained two final working samples of stars presented
in Figs. 4 and 5. The corresponding plots were constructed from a
sample of 45 000 red giant branch stars and 507 000 red clump
stars from the range of latitudes $2^\circ<|b|<10^\circ$. The mean
radius of the samples was $\overline {r}=5.1$~kpc for the red
giant branch stars and $\overline {r}=5.4$~kpc for the red clump
stars.

We hypothesize that the selected stars belong to the bar with a
fairly high probability, although there may be some fraction of
stars belonging to the bulge. Since the bulge consists mostly of
fairly old stars, we believe that the percentage of bulge stars in
the samples of red giant branch and red clump stars is low;
therefore, it is unlikely that they can have a noticeable effect
on the determination of bar characteristics.

\section*{RESULTS AND DISCUSSION}
In Fig. 4, the observed velocity components for the red giant
branch stars  $V_l=4.74 r \mu_l\cos b$ and $V_b=4.74 r \mu_b$ are
plotted against the Galactic longitude l. The expected velocity
dispersion (their value corresponds to the $3\sigma$ level) from
the model data by Zhao (1996) are marked in this figure (in the
form of check marks in a direction $l\approx0^\circ$); the gray
shading indicates the region that the Galactic circular orbits of
the stars occupy (the circular orbits for each stars were
calculated with parameters (8)). It can be seen from the figure
that the Galactic rotation has a significant effect when the
velocities $V_l$ are analyzed.

Figure 5 presents the distribution of stars in projection onto the
Galactic $XY$ plane. For ten directions along the line of sight,
we calculated the mean coordinates and their errors. It can be
seen from the figure that the part of the bar nearest to the Sun
is clearly revealed by both red giant branch and red clump stars.
Owing to the smaller number of stars in the sample and lesser
noisiness, the distribution of red giant branch stars seems more
illustrative. If we draw a line through points 3.8 (counted off
from left to right) in Fig. 5a (red giant branch stars), then the
bar inclination will be $35^\circ\pm10^\circ$ to the Galactic
center--Sun direction; it crosses the $OX$ axis at point
$X\approx7$~kpc, corresponding to $R_0=7$~kpc. In fact, however,
according to present-day measurements, $R_0=8.0$~kpc. If we draw a
line through points 3--4 and $X = 8$~kpc, then the inclination of
the bar axis will be $21^\circ\pm10^\circ$. Analysis of the
distribution of red clump stars (Fig. 5b) also yields similar
results, but, as can be seen from Fig. 5b, the mean values are
calculated from them with errors larger than those from red giant
branch stars by a factor of 1.5.

\begin{figure}[t]
{\begin{center}
 \includegraphics[width=140mm]{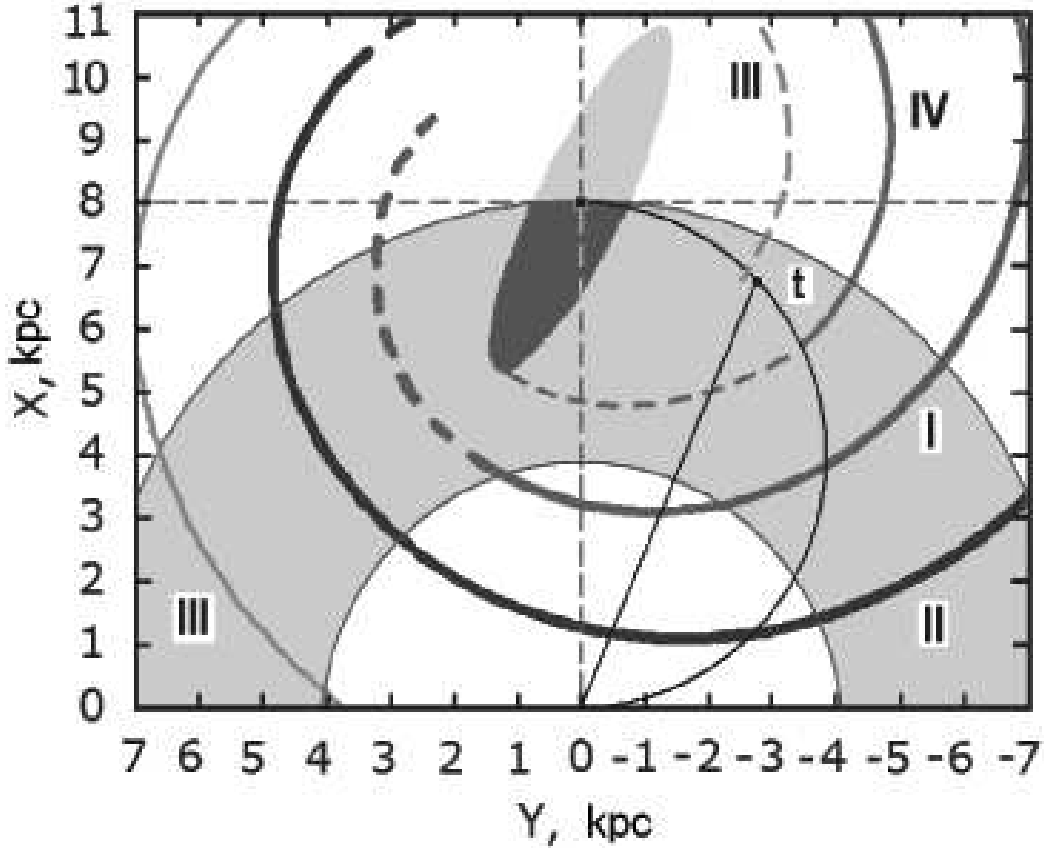}
 \caption{
Proposed model of the bar and refined (in the inner Galaxy) model
of a four-armed spiral pattern for the Galaxy, the gray shading in
the region 4 kpc $< r <$ 8 kpc indicates the model distribution of
stars, the light shading indicates the far, unobserved end of the
bar, the dark shading marks the observed region of the bar; for
additional explanations, see the text.}
 \label{f7}
\end{center}}
\end{figure}

In addition to the bar, the velocity peaks in Fig. 4 and the star
density maxima in Fig. 5 also reflect the spiral structure of the
Galaxy. An asymmetry is clearly seen in the distribution of stars
in Fig. 5, where there are much more stars in the right part of
the figure. The influence of the Carina–Sagittarius arm is In
addition to the bar, the velocity peaks in Fig. 4 and the star
density maxima in Fig. 5 also reflect the spiral structure of the
Galaxy. An asymmetry is clearly seen in the distribution of stars
in Fig. 5, where there are much more stars in the right part of
the figure. The influence of the Carina–Sagittarius arm is clearly
reflected in a direction $l\approx-15^\circ$, with this influence
being more tangible in the negative velocities $V_l$ (Fig. 4).

The solid lines in Fig. 6 indicate the four-armed spiral pattern
with the pitch angle $i = 13^\circ$ constructed from the data by
Bobylev and Bajkova (2013), in which the pattern parameters were
obtained by analyzing the distribution of maser sources with
measured trigonometric parallaxes in the Galaxy. The designations
of the spiral arms are as follows: I -- Scutum--Crux arm, II --
Carina--Sagittarius arm, III -- Perseus arm, IV -- Outer arm (or
Cygnus arm). The dashes indicate the extensions of the spiral arms
that, in our opinion, correspond more closely to the available
data. The thin black line indicates a semicircumference with the
radius $R_0/2=4$~kpc; the center of this circumference lies on the
Sun.Galactic center line. Such a circumference is the locus of
tangential points, the points where the circular rotation velocity
of the Galaxy lies exactly along the line of sight. Point $t$ seen
in the direction $l=-22^\circ$ is marked on the semicircumference.
According to radio-astronomical observations of the HI
line-of-sight velocities, it is in this direction that the 3-kpc
spiral arm is observed tangentially (Burton 1988). The distance
from the Galactic center to point t is 3 kpc (hence the name of
this arm). The figure shows the bar with a radius of 3 kpc
oriented at an angle of 25. to the Galactic center.Sun direction.
The near end of the bar is seen at the angle $l=10^\circ,$ which
well reflects our map of the distribution of stars (Fig. 5) and
the counts by Rattenbury et al. (2007).

The proposed scheme (Fig. 6) agrees well with the cartographic
model of a four-armed spiral pattern by Vall$\acute{e}$e (2008)
and the conclusion about the number of arms in the Galaxy reached
by Efremov (2011) by analyzing the large-scale distribution of
neutral, molecular, and ionized hydrogen clouds in the Galaxy.

Note that the spikes in $V_l$ and $V_b$ are observed when the line
of sight runs along the spiral arm. Note the significant clump of
stars in the left part of Fig. 5 at $l = 12^\circ$, which is
confirmed by the corresponding narrow peak in the velocity
distribution in Fig. 4. The stars in this direction are slightly
farther from the bar end (points 1.2); a similar zigzag at
$l=10^\circ$ is also seen in the estimates by Rattenbury et al.
(2007). This can be explained if the spiral arms form something in
the form of a ring or oval elongated along the major axis of the
bar. The line of sight will then run along this structure.

To better understand the situation in the immediate vicinity of
the bar, we performed Monte Carlo simulations. The results are
presented in Fig. 7. Figure 7a shows a map of the distribution of
stars corresponding to the model presented in Fig. 6. In the range
of distances $4 < r < 8$~kpc and in the spiral arm segments, the
stars were distributed uniformly randomly; in the bar, they were
distributed normally (we use only the near part of the bar,
because we do not see the far one). The sample consisted of 35 000
stars (brought closer to the sample of red giant branch stars).
One Monte Carlo realization obtained for model errors in the
stellar distances of 10\% is presented. Figure 7b shows a map of
the distribution of stars where a ring with a radius of 2 kpc
fringing the bar was added to the preceding model.

\begin{figure}[t]
{\begin{center}
 \includegraphics[width=140mm]{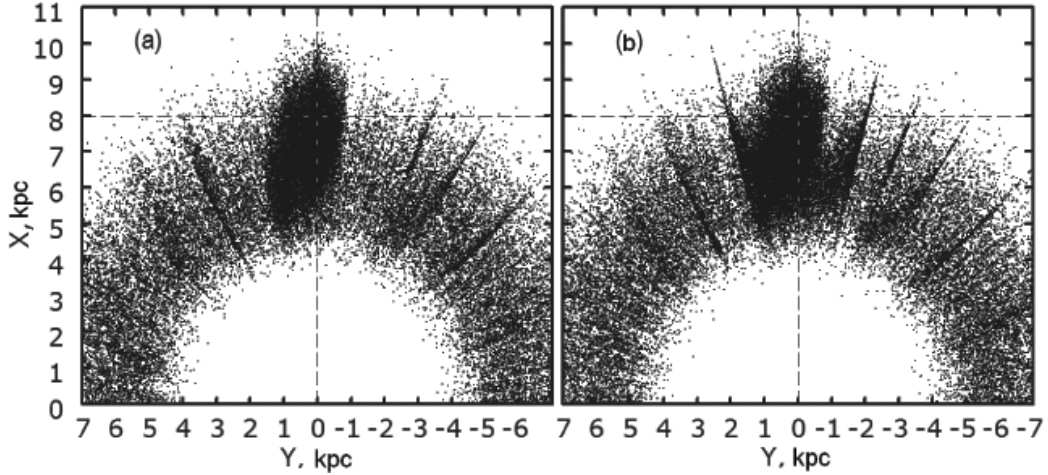}
 \caption{
Monte Carlo realization of the model presented in Fig. 6a, Monte
Carlo realization of the complicated model with the addition of a
ring 2 kpc in radius centered at the Galactic center; for
additional explanations, see the text. }
 \label{f8}
\end{center}}
\end{figure}

Our simulations lead us to several conclusions. (1) Although the
bar in the model is oriented at an angle of 25$^\circ$ to the
Galactic center -- Sun direction, the ellipse of the apparent
(smeared by the distance errors) distribution of stars is oriented
at a smaller angle of about 15$^\circ$. (2) In spite of the fact
that we used rather narrow spiral arms, our model data (Fig. 7)
well approximate the actually observed ones (Fig. 5). (3) The
model with a ring (Fig. 7b) provides a more accurate
approximation.

\section*{CONCLUSIONS}
We proposed and implemented a method for selecting stars belonging
to the Galactic bar. This method is based on 2MASS infrared
photometry and stellar proper motions, in particular, from the
Kharkiv XPM catalogue. It consists in the following: (1) the
pre-selection of red clump and red giant branch stars on the
color–magnitude diagram for which the photometric distances are
determined; (2) the elimination of background stars from the
preselected samples using constraints on the proper motions,
because the stellar proper motions are indicators of a larger
velocity dispersion toward the bar and the spiral arms compared to
the background stars. By the background we mean the stars with
circular orbits.

As a result, we obtained two samples of stars, 45 000 red giant
branch stars and 507 000 red clump stars from the range of
latitudes $2^\circ < |b| < 10^\circ$ at longitudes $|l| <
90^\circ$, based on which we mapped the distribution in the
Galactic XY plane.

Analysis of the maps allowed the bar segment nearest to the Sun
with an orientation of 20$^\circ$--25$^\circ$ with respect to the
Galactic bar -- Sun direction and a semimajor axis of no more than
3 kpc (short bar) to be identified with confidence.

Our numerical simulations of the velocities and distribution of
stars allowed the model of a four-armed spiral structure of the
Galaxy in the immediate vicinity of the bar to be refined. In
particular, we found arguments for the fact that an extension of
the Perseus arm(III) is the 3-kpc arm segment and that the Outer
arm (IV) begins from the bar end nearest to the Sun. It is quite
likely that all spiral arms merge into a 3-kpc oval (an ellipse or
ring). In fact, this may imply that two spiral arms can emerge
from each end of the bar.

\bigskip

\section*{ACKNOWLEDGMENTS}

We are grateful to the referee for valuable remarks that
contributed to a significant improvement of the paper. This work
was supported by the “Nonstationary Phenomena in Objects of the
Universe”Program of the Presidium of the Russian Academy of
Sciences, grant no. NSh-16245.2012.2 from the President of the
Russian Federation, and the Ministry of Education and Science of
the Russian Federation under contract no. 8417.

\bigskip

\bigskip

\noindent{\Large\bf REFERENCES}

\bigskip

\bigskip

1. C. Babusiaux and G. Gilmore, Mon. Not. R. Astron. Soc. 358,
1309 (2005).

2. R. A. Benjamin, E. Churchwell, B. L. Babler, et al., Astrophys.
J. 630, L149 (2005).

3. J. Binney, O. E. Gerhard, A. A. Stark, et al., Mon. Not. R.
Astron. Soc. 252, 210 (1991).

4. L. Blitz and D. N. Spergel, Astrophys. J. 379, 631 (1991).

5. V. V. Bobylev, A. T. Bajkova, and A. S. Stepanishchev, Astron.
Lett. 34, 515 (2008).

6. V. V. Bobylev and A. T. Bajkova, Astron. Lett. 39, 809 (2013).

7. W. B. Burton, {\it Galactic and Extragalactic Radio Astronomy},
Ed. by G. Verschuur and K. Kellerman (Springer, New York, 1988).

8. D. P. Clemens, Astrophys. J. 295, 422 (1985).

9. T. M. Dame and P. Thaddeus, Astron. J. 683, L143 (2008).

10. B. T. Draine, Ann. Rev. Astron. Astrophys. 41, 241 (2003).

11. Yu. N. Efremov, Astron.Rep. 55, 108 (2011).

12. P. Englmaier, M. Pohl, and N. Bissantz, Mem. Soc. Astron.
Ital. 18, 199 (2011).

13. P. N. Fedorov, A. A. Myznikov, and V. S. Akhmetov, Mon. Not.
R. Astron. Soc. 393, 133 (2009).

14. P. N. Fedorov, V. S. Akhmetov, V. V. Bobylev, and G. A.
Gontcharov, Mon. Not. R. Astron. Soc. 415, 665 (2011).

15. T. Foster and B. Cooper, ASP Conf. Ser. 438, 16 (2010).

16. C. Francis and E. Anderson, Mon. Not. R. Astron. Soc. 422,
1283 (2012).

17. G. A. Gontcharov, Astron. Lett. 37, 785 (2008).

18. G. A. Gontcharov, Astron. Lett. 37, 707 (2011).

19. E. Hog, C. Fabricius, V. V. Makarov, et al., Astron.
Astrophys. 355, L27 (2000).

20. L. G. Hou, J. L. Han, and W. B. Shi, Astron. Astrophys. 499,
473 (2009).

21. J. R. D. L$\acute{e}$pine, Yu. N. Mishurov, and S. Yu.Dedikov,
Astrophys. J. 546, 234 (2001).

22. M. L$\acute{o}$pez-Corredoira, A. Cabrera-Lavers, T. J.
Mahoney, et al., Astron. J. 133, 154 (2007).

23. D. J. Marshall, A. C. Robin, C. Reyle, et al., Astron.
Astrophys. 453, 635 (2006).

24. A. M. Mel’nik and P. Rautiainen, Astron. Lett. 35, 609 (2009).

25. Y. Nakada, T.Onaka, I. Yamamura, et al., Nature 353, 140
(1991).

26. D. M. Nataf, A. Gould, P. Fouqu, et al., Astrophys. J. 769, 88
(2013).

27. N. J. Rattenbury, S. Mao, T. Sumi, et al., Mon. Not. R.
Astron. Soc. 378, 1064 (2007).

28. M. Reid and A. Brunthaler, Astrophys. J. 616, 872 (2004).

29. D. Russeil, Astron. Astrophys. 397, 133 (2003).

30. A. Sanna,M. J. Reid, L. Moscadelli, et al., Astrophys. J. 706,
464 (2009).

31. M. F. Skrutskie, R. M. Cutri, R. Stiening, et al., Astron. J.
131, 1163 (2006).

32. K.Z.Stanek,M.Mateo, A. Udalski, et al.,Astrophys. J. 429, L73
(1994).

33. A. Udalski, M. Szymanski, K. Z. Stanek, et al., Acta Astron.
44, 165 (1994).

34. J. P. Vall$\acute{e}$e, Astron. J. 135, 1301 (2008).

35. G. de Vaucouleurs, in {\it The Spiral Structure of Our
Galaxy}, Ed. W. Becker and G. Contopoulos (Reidel, Dordrecht,
1970), p. 18.

36. Y.Wang,H. Zhao, and S. Mao,Mon. Not.R. Astron. Soc. 427, 1429
(2012).

37. H. Zhao,Mon. Not. R. Astron. Soc. 283, 149 (1996).

\end{document}